\documentclass[11pt]{article}
\usepackage[utf8]{inputenc}

\usepackage{tikz}
\usetikzlibrary{positioning}
\usetikzlibrary{fit}

\usepackage{adjustbox}

\usepackage{amsmath}
\usepackage{amssymb}
\usepackage{amsthm}
\usepackage{authblk}

\usepackage{algpseudocode}
\usepackage{algorithm}
\usepackage{tikz}
\usetikzlibrary{positioning}

\usepackage{enumitem}

\usepackage{fullpage}

\usepackage{xcolor}

\newenvironment{myproof}[1][\proofname]{
  \begin{proof}[#1]
}{
  \end{proof}
}
\usepackage{hyperref}
\usepackage[all]{hypcap}

\title{A Critique of Chen's ``The 2-MAXSAT Problem Can Be Solved in Polynomial Time''\thanks{Supported in part by NSF grant 
		CCF-2006496.}}

\author{Tran~Duy~Anh~Le}
\author{Michael~P.~Reidy}
\author{Eliot~J.~Smith}
\affil{Department of Computer Science\\University of Rochester\\Rochester, NY 14627, USA}

\newcommand{\p}{\ensuremath{{\rm P}}}
\newcommand{\np}{\ensuremath{{\rm NP}}}

\newtheorem{example}{Example}
\newenvironment{examplecont}
{\addtocounter{example}{-1}\begin{example}{\textit{\textbf{{(continued)}}}}}
  {\end{example}}

\newtheorem{theorem}{Theorem}
\newtheorem{proposition}[theorem]{Proposition}

\date{February 21, 2024}

\begin{document}\sloppy

\maketitle

\begin{abstract}

In this paper, we examine Yangjun Chen's technical report titled ``The 2-MAXSAT Problem Can Be Solved in Polynomial Time''~\cite{chen:2-maxsat-solved-in-p-time}, which revises and expands upon their conference paper of the same name~\cite{chen:2-maxsat-solved-in-p-time-conf}.
Chen's paper purports to build a polynomial-time algorithm for the NP-complete problem 2-MAXSAT by converting a 2-CNF formula into a graph that is then searched.
We show through multiple counterexamples that Chen's proposed algorithms contain flaws, and we find that the structures they create lack properly formalized definitions. Furthermore, we elaborate on how the author fails to prove the correctness of their algorithms and how they make overgeneralizations in their time analysis of their proposed solution. Due to these issues, we conclude that Chen's technical report~\cite{chen:2-maxsat-solved-in-p-time} and conference paper~\cite{chen:2-maxsat-solved-in-p-time-conf} both fail to provide a proof that $\p=\np$.

\end{abstract}

\section{Introduction}

This critique looks at Yangjun Chen's paper titled ``The 2-MAXSAT Problem Can Be Solved in Polynomial Time'' \cite{chen:2-maxsat-solved-in-p-time}, which claims to give a polynomial-time algorithm for solving the well-known $\np$-complete problem 2-MAXSAT~\cite{gar-joh-sto:simplified-np-complete-problems}(also known as maximum~2-satisfiability or MAX~\mbox{2-SAT}). In their paper, Chen claims to provide a proof that such an algorithm exists, which would also be a proof that $\p = \np$.

In this paper, we argue that the algorithm for solving the 2-MAXSAT problem given by Chen does not solve the 2-MAXSAT problem by providing cases where Chen's algorithm produces incorrect results. Additionally, we remark that due to poor algorithm formalizations, arguments that lack support, and overgeneralizations, it is possible that Chen's algorithm may not run in polynomial-time.

Our Section~\ref{s:prelims} covers preliminary topics related to our paper. Section~\ref{s:overview} gives a high-level overview of Chen's proposed solution, including an example. Section~\ref{s:search-find-subset} examines Chen's Algorithm~1 (also called their \textit{SEARCH} algorithm) and Algorithm~2 (also called their \textit{findSubset} algorithm). For the former we give examples that cause the algorithm to fail, and for the latter we cast doubt on the validity of Chen's proof of a polynomial runtime.
Section~\ref{s:graph-improvements} examines Chen's graph improvements and shows an example where their Algorithm~3 (which is claimed to be an improved version of their Algorithm~1) fails. Section~\ref{s:complexity} examines Chen's analysis of the overall complexity of their solution and finds several flaws. Finally, Section~\ref{s:examples-over-proofs} notes Chen's problematic use of examples over proofs in attempting to show correctness.

We critique here the technical report version~\cite{chen:2-maxsat-solved-in-p-time} of Chen's paper. However, their paper also appears as a conference version~\cite{chen:2-maxsat-solved-in-p-time-conf}. We chose to critique Chen's technical report since it is the most detailed and up-to-date version of their paper, and it claims a looser polynomial-time bound for the runtime of their algorithm than the bound that is claimed in the conference version, which suggests that the bound claimed in their conference version may be incorrect. Furthermore, the analysis section from the conference version~\cite{chen:2-maxsat-solved-in-p-time-conf} is completely removed from their technical report~\cite{chen:2-maxsat-solved-in-p-time}, and replaced with a new analysis of an algorithm that Chen claims is an improvement upon that of the conference version's. However, the fact that the conference version's analysis was entirely removed leads us to doubt it's correctness. The conference version covers a subset of the sections contained in the technical report, including Sections 1--3 and Algorithms~1 and 2, and the technical report builds upon the conference version with an updated improvements section. Any of our findings that apply to these sections and algorithms, found in Sections \ref{s:search-find-subset} and \ref{s:examples-over-proofs} of this paper, apply to the conference version of Chen's paper as well; Sections~\ref{s:graph-improvements} and \ref{s:complexity} of this paper critique content appearing only in Chen's technical report and are not applicable to their conference version. However, the content we provide that is applicable to Chen's conference version~\cite{chen:2-maxsat-solved-in-p-time-conf} is more than enough to disprove that the algorithm detailed in that paper is not able to solve 2-MAXSAT.

\section{Preliminaries}
\label{s:prelims}

In this paper, we assume the reader has a foundational understanding of algorithmic complexity analysis, the P, NP, and \mbox{NP-complete} complexity classes, graph structures, trie structures, and set operations. We also assume basic knowledge of boolean algebra, boolean formulas, and the SAT problem.

Occasionally, we deviate from the original notation of Chen's paper to ensure clearer comprehension. When these changes occur, they will be pointed out alongside the original notation.

\subsection{Types of Boolean Formulas}

In this paper we refer to boolean formulas of two specific forms: 2-conjunctive normal form (2-CNF) and 2-disjunctive normal form (2-DNF). A boolean formula in 2-CNF is a conjunction of clauses, where each clause is the disjunction of two literals (a literal being either a variable or a negated variable), e.g.,\ $(v_1 \vee \neg v_2) \wedge (v_2 \vee v_3) \wedge (\neg v_1 \vee v_3)$. A boolean formula in 2-DNF is a disjunction of conjunctions, where each conjunction is between two literals, e.g., $(v_1 \wedge \neg v_1) \vee (v_2 \wedge v_3) \vee (\neg v_1 \wedge v_2)$.

\subsection{The 2-MAXSAT Problem}

Informally, given a boolean formula $f$ in 2-CNF, the 2-MAXSAT problem asks if at least $k$ clauses in $f$ can be satisfied under a single assignment to the variables in $f$. Previous work has shown that 2-MAXSAT is an $\np$-complete problem through a polynomial-time many-one reduction from 3-SAT~\cite{gar-joh-sto:simplified-np-complete-problems}. We define the 2-MAXSAT problem as follows~\cite{chen:2-maxsat-solved-in-p-time}.

\begin{itemize}[label={}]
\item \textbf{Name:} 2 maximum satisfiability (2-MAXSAT).

\item \textbf{Given:} A set of boolean variables {V}, a boolean formula $f$ in 2-CNF form that is the conjunction of the clauses $C_1, C_2, \ldots, C_n$ with each clause containing at most $2$ disjunctive literals from $V$, and a positive integer $k < n$.

\item \textbf{Question:} Is there a truth assignment to $V$ that will satisfy at least $k$ clauses in $f$?
\end{itemize}

When we speak of the ``maximum number of satisfied clauses in a formula $F$,'' what we always mean is the number of clauses in $F$ that simultaneously evaluate to true under the variable assignment to $F$ (from all possible variables assignments) that maximizes this number. 

\section{Overview of Chen's Solution}
\label{s:overview}

We detail here, informally, a high-level step-by-step overview of Chen's solution to the \mbox{2-MAXSAT} problem. Below the description of each step, we build upon a running example.
In later sections, when simulating Chen's algorithms applied to example 2-CNF formulas, we omit the details of more trivial steps for the sake of brevity.

\begin{description}
\item{} Begin with a 2-CNF formula over the set of boolean variables $V$, call it $F$, of the form $c_1 \wedge c_2 \wedge \cdots \wedge c_n$, where each $c_i$ is a clause and there are $n$ clauses.
\begin{example}
    $F = (v_1 \vee \neg v_2) \wedge (\neg v_1 \vee v_3).$
\end{example}
\item{Step 1.} Convert $F$ to a 2-DNF formula, call it $D$, of the form $d_1 \vee d_2 \vee \cdots \vee d_n$, where each clause $c_i = (l_1 \vee l_2)$ is turned into two conjunctions of the form $d_i = (l_1 \wedge y_i) \vee (l_2 \wedge \neg y_i)$ ($y_i$ is a new variable that does not appear in $V$).
\begin{examplecont}
    $D = \underbrace{(v_1 \wedge y_1)}_ \text{Conjunction a} \vee \underbrace{(\neg v_2 \wedge \neg y_1)}_ \text{Conjunction b} \vee \underbrace{(\neg v_1 \wedge y_2)}_ \text{Conjunction c} \vee \underbrace{(v_3 \wedge \neg y_2)}_ \text{Conjunction d}.$
\end{examplecont}
\item{Step 2.} Create a boolean formula $D'$ that is a modification of $D$. $D'$ is initially identical to $D$, except that if a variable $v_i$ that appears in $D$ is absent from a conjunction $d$ in $D$, the clause $(v_i \vee \neg v_i)$ is added to $d$ (appended with $\wedge$). Chen uses a special syntax $(v_i,*)$ to represent these clauses, and we will also refer to variables such as $v_i$ as ``missing'' variables.
\begin{examplecont}
    $\begin{aligned}[t] D' =~&\underbrace{\bigl(v_1 \wedge y_1 \wedge (v_2,*) \wedge (v_3,*) \wedge (y_2,*)\bigr)}_ \text{a} \vee ~
    \\
    &\underbrace{\bigl(\neg v_2 \wedge \neg y_1 \wedge (v_1,*) \wedge (v_3,*) \wedge (y_2,*)\bigr)}_ \text{b} \vee ~
    \\
    &\underbrace{\bigl(\neg v_1 \wedge y_2 \wedge (v_2,*) \wedge (v_3,*) \wedge (y_1,*)\bigr)}_ \text{c} \vee ~
    \\
    &\underbrace{\bigl(v_3 \wedge \neg y_2 \wedge (v_1,*) \wedge (v_2,*) \wedge (y_1,*)\bigr)}_ \text{d}.\end{aligned}$
\end{examplecont}
\item{Step 3.} Convert each conjunction in $D'$ into a sequence of variables with all negated variables removed, of the form $v_1.v_2.\cdots.v_k$, where each $v_i$ is either a variable or a missing variable from the conjunction. Matching Chen's syntax, the `.' delineate the elements of the sequence.
\begin{examplecont}
    Unsorted variable sequences.
    \begin{align}
        &v_1.y_1.(v_2,*).(v_3,*).(y_2,*) \tag{a}
        \\
        &(v_1,*).(v_3,*).(y_2,*) \tag{b}
        \\
        &y_2.(v_2,*).(v_3,*).(y_1,*) \tag{c}
        \\
        &v_3.(v_1,*).(v_2,*).(y_1,*) \tag{d}
    \end{align}
\end{examplecont}
\item{Step 4.} Sort the sequences of variables using any global ordering over the variables (Chen uses the frequency of the variables across all of the sequences to build the ordering~\cite{chen:2-maxsat-solved-in-p-time}). The global ordering is fixed for the algorithm, i.e.\ the ordering is not passed as input to the algorithm. After the sorting, add the start character `\#' to the front of each sequence and the terminal character `\$' to the end of each sequence.
\begin{examplecont}
    Variable sequences sorted using a lexical global ordering.
    \begin{align}
        &\#.v_1.(v_2,*).(v_3,*).y_1.(y_2,*).\$ \tag{a}
        \\
        &\#.(v_1,*).(v_3,*).(y_2,*).\$ \tag{b}
        \\
        &\#.(v_2,*).(v_3,*).(y_1,*).y_2.\$ \tag{c}
        \\
        &\#.(v_1,*).(v_2,*).v_3.(y_1,*).\$ \tag{d}
    \end{align}
\end{examplecont}
\item{Step 5.} Turn each sorted sequence into a $p$-graph, which is a graph with nodes corresponding to the variables of the sequence (in order) and edges connecting nodes that represent adjacent variables in the sequence. A $p$-graph also has dashed edges that Chen calls ``spans''~\cite{chen:2-maxsat-solved-in-p-time}. A span is an edge that ``jumps over'' a missing variable (of the form $(v_k,*)$) to connect the variables before and after the missing variable, represented as $\langle v_i, v_k, v_j \rangle$ where $v_k$ is the missing variable. Chen occasionally calls $p$-graphs ``$p$-paths''~\cite{chen:2-maxsat-solved-in-p-time}, however we refer to them only as $p$-graphs.
\begin{examplecont}
    See Figure~\ref{fig:example-p-graphs} for the $p$-graphs corresponding to the four sorted sequences.
    \begin{figure}
        \centering    
        \begin{adjustbox}{height=250pt}
        \begin{tikzpicture}
            [rnode/.style={circle, draw=black, thick, minimum size=7mm},]
    
            \node[rnode] (c1) [] {\#};
            \node[rnode] (c2) [below=of c1] {$v_1$};
            \node[rnode] (c3) [below=of c2] {$v_2$};
            \node[rnode] (c4) [below=of c3] {$v_3$};
            \node[rnode] (c5) [below=of c4] {$y_1$};
            \node[rnode] (c6) [below=of c5] {$y_2$};
            \node[rnode] (c11) [below=of c6, label=below:{a}] {\$};

            \node[rnode, draw=none] (d1) [right=of c1] {};
            
            \node[rnode] (c7) [right=of d1] {\#};
            \node[rnode] (c8) [below=of c7] {$v_1$};
            \node[rnode] (c9) [below=of c8] {$v_3$};
            \node[rnode] (c10) [below=of c9] {$y_2$};
            \node[rnode] (c12) [below=of c10, label=below:{b}] {\$};

            \node[rnode, draw=none] (d2) [right=of c7] {};
    
            \node[rnode] (c13) [right=of d2] {\#};
            \node[rnode] (c14) [below=of c13] {$v_2$};
            \node[rnode] (c15) [below=of c14] {$v_3$};
            \node[rnode] (c16) [below=of c15] {$y_1$};
            \node[rnode] (c17) [below=of c16] {$y_2$};
            \node[rnode] (c18) [below=of c17, label=below:{c}] {\$};

            \node[rnode, draw=none] (d3) [right=of c13] {};
    
            \node[rnode] (c19) [right=of d3] {\#};
            \node[rnode] (c20) [below=of c19] {$v_1$};
            \node[rnode] (c21) [below=of c20] {$v_2$};
            \node[rnode] (c22) [below=of c21] {$v_3$};
            \node[rnode] (c23) [below=of c22] {$y_1$};
            \node[rnode] (c24) [below=of c23, label=below:{d}] {\$};
            
            \draw[-] (c1.south) -- (c2.north);
            \draw[-] (c2.south) -- (c3.north);
            \draw[-] (c3.south) -- (c4.north);
            \draw[-] (c4.south) -- (c5.north);
            \draw[-] (c5.south) -- (c6.north);
            \draw[-] (c6.south) -- (c11.north);
    
            \draw[-] (c7.south) -- (c8.north);
            \draw[-] (c8.south) -- (c9.north);
            \draw[-] (c9.south) -- (c10.north);
            \draw[-] (c10.south) -- (c12.north);
    
            \draw[-] (c13.south) -- (c14.north);
            \draw[-] (c14.south) -- (c15.north);
            \draw[-] (c15.south) -- (c16.north);
            \draw[-] (c16.south) -- (c17.north);
            \draw[-] (c17.south) -- (c18.north);
            
            \draw[-] (c19.south) -- (c20.north);
            \draw[-] (c20.south) -- (c21.north);
            \draw[-] (c21.south) -- (c22.north);
            \draw[-] (c22.south) -- (c23.north);
            \draw[-] (c23.south) -- (c24.north);

            \draw[dashed] (c4.north west) to[out=110,in=250] (c2.south west);
            \draw[dashed] (c5.north west) to[out=110,in=250] (c3.south west);
            \draw[dashed] (c11.north west) to[out=110,in=250] (c5.south west);

            \draw[dashed] (c9.north west) to[out=110,in=250] (c7.south west);
            \draw[dashed] (c10.north west) to[out=110,in=250] (c8.south west);
            \draw[dashed] (c12.north west) to[out=110,in=250] (c9.south west);

            \draw[dashed] (c15.north west) to[out=110,in=250] (c13.south west);
            \draw[dashed] (c16.north west) to[out=110,in=250] (c14.south west);
            \draw[dashed] (c17.north west) to[out=110,in=250] (c15.south west);

            \draw[dashed] (c21.north west) to[out=110,in=250] (c19.south west);
            \draw[dashed] (c22.north west) to[out=110,in=250] (c20.south west);
            \draw[dashed] (c24.north west) to[out=110,in=250] (c22.south west);
        \end{tikzpicture}
        \end{adjustbox}
        \caption{The $p$-graphs for the running example.} \label{fig:example-p-graphs}
    \end{figure}
\end{examplecont}
\item{Step 6.} Convert each $p$-graph into a $p$*-graph (which we create now but use later in Step~8) by merging the spans of the $p$-graph that overlap to create larger spans: For two overlapping spans $\langle v_i,\cdots,v_k,v_j \rangle$ and $\langle v_k,v_j,\cdots,v_l \rangle$, i.e.\ the two-node suffix of the first span is identical to the two-node prefix of the second span, a new span $\langle v_i,\cdots,v_k,v_j,\cdots,v_l \rangle$ is added to the $p$*-graph (without removing the original two spans). These successively larger spans are added to the $p$*-graph until all overlapping spans have been merged. Chen calls this merging process the ``transitive closure''~\cite{chen:2-maxsat-solved-in-p-time} of the base spans (the spans of the $p$-graph, consisting of three variables).
\begin{examplecont}
    See Figure~\ref{fig:example-p*-graphs} for the $p$*-graphs.
    \begin{figure}
        \centering
        \begin{adjustbox}{height=250pt}
        \begin{tikzpicture}
            [rnode/.style={circle, draw=black, thick, minimum size=7mm},]
    
            \node[rnode] (c1) [] {\#};
            \node[rnode] (c2) [below=of c1] {$v_1$};
            \node[rnode] (c3) [below=of c2] {$v_2$};
            \node[rnode] (c4) [below=of c3] {$v_3$};
            \node[rnode] (c5) [below=of c4] {$y_1$};
            \node[rnode] (c6) [below=of c5] {$y_2$};
            \node[rnode] (c11) [below=of c6, label=below:{a}] {\$};

            \node[rnode, draw=none] (d1) [right=of c1] {};
            
            \node[rnode] (c7) [right=of d1] {\#};
            \node[rnode] (c8) [below=of c7] {$v_1$};
            \node[rnode] (c9) [below=of c8] {$v_3$};
            \node[rnode] (c10) [below=of c9] {$y_2$};
            \node[rnode] (c12) [below=of c10, label=below:{b}] {\$};

            \node[rnode, draw=none] (d2) [right=of c7] {};
    
            \node[rnode] (c13) [right=of d2] {\#};
            \node[rnode] (c14) [below=of c13] {$v_2$};
            \node[rnode] (c15) [below=of c14] {$v_3$};
            \node[rnode] (c16) [below=of c15] {$y_1$};
            \node[rnode] (c17) [below=of c16] {$y_2$};
            \node[rnode] (c18) [below=of c17, label=below:{c}] {\$};

            \node[rnode, draw=none] (d3) [right=of c13] {};
    
            \node[rnode] (c19) [right=of d3] {\#};
            \node[rnode] (c20) [below=of c19] {$v_1$};
            \node[rnode] (c21) [below=of c20] {$v_2$};
            \node[rnode] (c22) [below=of c21] {$v_3$};
            \node[rnode] (c23) [below=of c22] {$y_1$};
            \node[rnode] (c24) [below=of c23, label=below:{d}] {\$};
            
            \draw[-] (c1.south) -- (c2.north);
            \draw[-] (c2.south) -- (c3.north);
            \draw[-] (c3.south) -- (c4.north);
            \draw[-] (c4.south) -- (c5.north);
            \draw[-] (c5.south) -- (c6.north);
            \draw[-] (c6.south) -- (c11.north);
    
            \draw[-] (c7.south) -- (c8.north);
            \draw[-] (c8.south) -- (c9.north);
            \draw[-] (c9.south) -- (c10.north);
            \draw[-] (c10.south) -- (c12.north);
    
            \draw[-] (c13.south) -- (c14.north);
            \draw[-] (c14.south) -- (c15.north);
            \draw[-] (c15.south) -- (c16.north);
            \draw[-] (c16.south) -- (c17.north);
            \draw[-] (c17.south) -- (c18.north);
            
            \draw[-] (c19.south) -- (c20.north);
            \draw[-] (c20.south) -- (c21.north);
            \draw[-] (c21.south) -- (c22.north);
            \draw[-] (c22.south) -- (c23.north);
            \draw[-] (c23.south) -- (c24.north);
            
            \draw[dashed] (c4.north west) to[out=110,in=250] (c2.south west);
            \draw[dashed] (c5.north west) to[out=110,in=250] (c3.south west);
            \draw[dashed] (c11.north west) to[out=110,in=250] (c5.south west);
            \draw[dashed] (c5.north west) to[out=120,in=240] (c2.south west);

            \draw[dashed] (c12.north west) to[out=110,in=250] (c9.south west);
            \draw[dashed] (c10.north west) to[out=110,in=250] (c8.south west);
            \draw[dashed] (c9.north west) to[out=110,in=250] (c7.south west);
            \draw[dashed] (c12.north west) to[out=120,in=240] (c8.south west);
            \draw[dashed] (c10.north west) to[out=120,in=240] (c7.south west);
            \draw[dashed] (c12.north west) to[out=130,in=230] (c7.south west);

            \draw[dashed] (c17.north west) to[out=110,in=250] (c15.south west);
            \draw[dashed] (c16.north west) to[out=110,in=250] (c14.south west);
            \draw[dashed] (c15.north west) to[out=110,in=250] (c13.south west);
            \draw[dashed] (c17.north west) to[out=120,in=240] (c14.south west);
            \draw[dashed] (c16.north west) to[out=120,in=240] (c13.south west);
            \draw[dashed] (c17.north west) to[out=130,in=230] (c13.south west);

            \draw[dashed] (c24.north west) to[out=110,in=250] (c22.south west);
            \draw[dashed] (c21.north west) to[out=110,in=250] (c19.south west);
            \draw[dashed] (c22.north west) to[out=110,in=250] (c20.south west);
            \draw[dashed] (c22.north west) to[out=120,in=240] (c19.south west);
        \end{tikzpicture}
        \end{adjustbox}
        \caption{The $p$*-graphs for the running example.} \label{fig:example-p*-graphs}
    \end{figure}
\end{examplecont}
\item{Step 7.} Independent of Step~6, merge what Chen calls the ``main paths''~\cite{chen:2-maxsat-solved-in-p-time} of the $p$-graphs into a single trie, call it $T$, with the merging process working as follows: For the set of $p$-graphs $R$, divide $R$ into subsets where, within each subset, the $p$-graphs have the same label on their first node (first with regards to the sequence the $p$-graph is derived from, or uppermost with regards to the $p$-graph itself). Within a subset, call it $R_i$, remove the first node, labeled with $v$, from each $p$-graph. Connect all one-node-smaller $p$-graphs with only one node (which must be labelled with \$) to a new root node labeled with $v$, and label the only node to indicate which $p$-graph that trie branch corresponds to. Recursively merge the remaining one-node-smaller $p$-graphs with two or more nodes, resulting in a set of tries, and connect the root of those tries to the new root. The new root should have some metadata such that all of the removed first nodes of the $p$-graphs (with the same label) now correspond to the new root (Chen does not mention or detail this metadata). During this step, the spans of the $p$-graphs are ignored.
\begin{examplecont}
    See Figure~\ref{fig:example-trie} for the trie $T$ produced by the recursive merging process.
    \begin{figure}
        \centering
        \begin{adjustbox}{height=280pt}
        \begin{tikzpicture}
            [rnode/.style={circle, draw=black, thick, minimum size=7mm},]
    
            \node[rnode] (c1) [label=above left:{$n_1$}] {\#};
            \node[rnode] (c2) [below left=of c1, label=above left:{$n_2$}] {$v_1$};
            \node[rnode] (c3) [below=of c2, label=right:{$n_6$}] {$v_2$};
            \node[rnode] (c4) [below=of c3, label=right:{$n_7$}] {$v_3$};
            \node[rnode] (c5) [below=of c4, label=right:{$n_8$}] {$y_1$};
            \node[rnode] (c6) [below=of c5, label=right:{$n_9$}] {$y_2$};
            \node[rnode] (c11) [below=of c6, label=below:{$\{a\}$}, label=right:{$n_{10}$}] {\$};
            
            \node[rnode] (c9) [left=of c3, label=above left:{$n_3$}] {$v_3$};
            \node[rnode] (c10) [below=of c9, label=left:{$n_4$}] {$y_2$};
            \node[rnode] (c12) [below=of c10, label=below:{$\{b\}$}, label=left:{$n_5$}] {\$};

            \node[rnode, draw=none] (d1) [] {};
    
            \node[rnode] (c14) [below right=of c1, label=above right:{$n_{12}$}] {$v_2$};
            \node[rnode] (c15) [below=of c14, label=right:{$n_{13}$}] {$v_3$};
            \node[rnode] (c16) [below=of c15, label=right:{$n_{14}$}] {$y_1$};
            \node[rnode] (c17) [below=of c16, label=right:{$n_{15}$}] {$y_2$};
            \node[rnode] (c18) [below=of c17, label=below:{$\{c\}$}, label=right:{$n_{16}$}] {\$};

            \node[rnode] (c24) [right=of c6, label=below:{$\{d\}$}, label=above:{$n_{11}$}] {\$};
            
            \draw[-] (c1.south west) -- (c2.north east);
            \draw[-] (c2.south) -- (c3.north);
            \draw[-] (c3.south) -- (c4.north);
            \draw[-] (c4.south) -- (c5.north);
            \draw[-] (c5.south) -- (c6.north);
            \draw[-] (c6.south) -- (c11.north);
    
            \draw[-] (c2.south west) -- (c9.north east);
            \draw[-] (c9.south) -- (c10.north);
            \draw[-] (c10.south) -- (c12.north);
    
            \draw[-] (c1.south east) -- (c14.north west);
            \draw[-] (c14.south) -- (c15.north);
            \draw[-] (c15.south) -- (c16.north);
            \draw[-] (c16.south) -- (c17.north);
            \draw[-] (c17.south) -- (c18.north);

            \draw[-] (c5.south east) -- (c24.north west);
        \end{tikzpicture}
        \end{adjustbox}
        \caption{The trie for the running example.} \label{fig:example-trie}
    \end{figure}
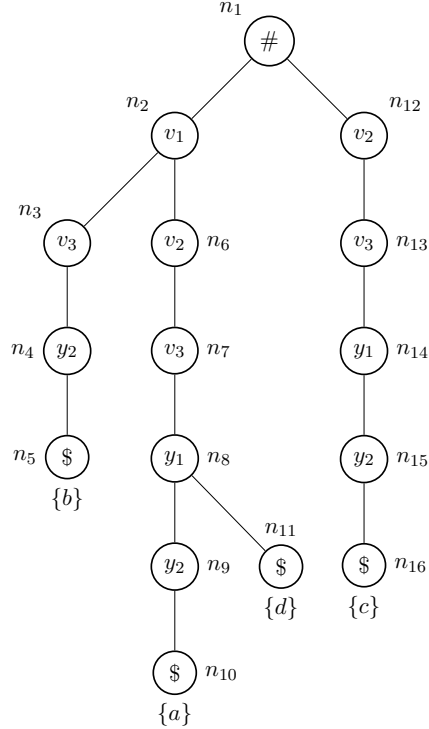
\end{examplecont}
\item{Step 8.} Convert the trie $T$ into a trie-like graph $G$ by adding to $T$ the spans of each $p$*-graph (which can result in cycles and thus makes $T$ no longer a true trie). Chen does not specify how to add the spans to $T$~\cite{chen:2-maxsat-solved-in-p-time}, however we presume that, as touched on in Step 7, there is some metadata that establishes a correspondence between nodes of the $p$-graphs/$p$*-graphs and the nodes of $T$.
\begin{examplecont}
    See Figure~\ref{fig:example-trie-like-graph} for the trie-like graph $G$\@. While Chen does not algorithmically describe how to create the trie-like graph, $G$ was constructed based on Chen's informal descriptions of the trie-like graphs and the process of creating them~\cite{chen:2-maxsat-solved-in-p-time}.
    \begin{figure}
        \centering
        \begin{adjustbox}{height=280pt}
        \begin{tikzpicture}
            [rnode/.style={circle, draw=black, thick, minimum size=7mm},]
    
            \node[rnode] (c1) [label=above left:{$n_{1}$}] {\#};
            \node[rnode] (c2) [below left=of c1, label=above left:{$n_{2}$}] {$v_1$};
            \node[rnode] (c3) [below=of c2, label=right:{$n_{6}$}] {$v_2$};
            \node[rnode] (c4) [below=of c3, label=right:{$n_{7}$}] {$v_3$};
            \node[rnode] (c5) [below=of c4, label=right:{$n_{8}$}] {$y_1$};
            \node[rnode] (c6) [below=of c5, label=right:{$n_{9}$}] {$y_2$};
            \node[rnode] (c11) [below=of c6, label=below:{$\{a\}$}, label=right:{$n_{10}$}] {\$};
            
            \node[rnode] (c9) [left=of c3, label=right:{$n_{3}$}] {$v_3$};
            \node[rnode] (c10) [below=of c9, label=right:{$n_{4}$}] {$y_2$};
            \node[rnode] (c12) [below=of c10, label=below:{$\{b\}$}, label=right:{$n_{5}$}] {\$};

            \node[rnode, draw=none] (d1) [] {};
    
            \node[rnode] (c14) [below right=of c1, label=above:{$n_{12}$}] {$v_2$};
            \node[rnode] (c15) [below=of c14, label=right:{$n_{13}$}] {$v_3$};
            \node[rnode] (c16) [below=of c15, label=right:{$n_{14}$}] {$y_1$};
            \node[rnode] (c17) [below=of c16, label=right:{$n_{15}$}] {$y_2$};
            \node[rnode] (c18) [below=of c17, label=below:{$\{c\}$}, label=right:{$n_{16}$}] {\$};

            \node[rnode] (c24) [right=of c6, label=below:{$\{d\}$}, label=above right:{$n_{11}$}] {\$};
            
            \draw[-] (c1.south west) -- (c2.north east);
            \draw[-] (c2.south) -- (c3.north);
            \draw[-] (c3.south) -- (c4.north);
            \draw[-] (c4.south) -- (c5.north);
            \draw[-] (c5.south) -- (c6.north);
            \draw[-] (c6.south) -- (c11.north);
    
            \draw[-] (c2.south west) -- (c9.north east);
            \draw[-] (c9.south) -- (c10.north);
            \draw[-] (c10.south) -- (c12.north);
    
            \draw[-] (c1.south east) -- (c14.north west);
            \draw[-] (c14.south) -- (c15.north);
            \draw[-] (c15.south) -- (c16.north);
            \draw[-] (c16.south) -- (c17.north);
            \draw[-] (c17.south) -- (c18.north);

            \draw[-] (c5.south east) -- (c24.north west);

            \draw[dashed] (c4.north west) to[out=110,in=250] (c2.south west);
            \draw[dashed] (c5.north west) to[out=110,in=250] (c3.south west);
            \draw[dashed] (c11.north west) to[out=110,in=250] (c5.south west);
            \draw[dashed] (c5.north west) to[out=120,in=240] (c2.south west);

            \draw[dashed] (c12.north west) to[out=110,in=250] (c9.south west);
            \draw[dashed, looseness=1.2] (c10.north west) to[out=110,in=180] (c2.west);
            \draw[dashed] (c9.north) to[out=90,in=180] (c1.west);
            \draw[dashed, looseness=1.4] (c12.north west) to[out=120,in=180] (c2.west);
            \draw[dashed, looseness=1.4] (c10.north west) to[out=120,in=180] (c1.west);
            \draw[dashed, looseness=1.6] (c12.north west) to[out=130,in=180] (c1.west);

            \draw[dashed] (c17.north west) to[out=110,in=250] (c15.south west);
            \draw[dashed] (c16.north west) to[out=110,in=250] (c14.south west);
            \draw[dashed, looseness=1.2] (c15.north east) to[out=50,in=0] (c1.east);
            \draw[dashed] (c17.north west) to[out=120,in=240] (c14.south west);
            \draw[dashed, looseness=1.2] (c16.north east) to[out=40,in=0] (c1.east);
            \draw[dashed, looseness=1.2] (c17.north east) to[out=50,in=0] (c1.east);

            \draw[dashed] (c3.north east) to[out=40,in=270] (c1.south);
            \draw[dashed] (c4.north east) to[out=50,in=270] (c1.south);
            \draw[dashed] (c24.north) to[out=90,in=-45] (c4.south east);
        \end{tikzpicture}
        \end{adjustbox}
        \caption{The trie-like graph for the running example.} \label{fig:example-trie-like-graph}
    \end{figure}
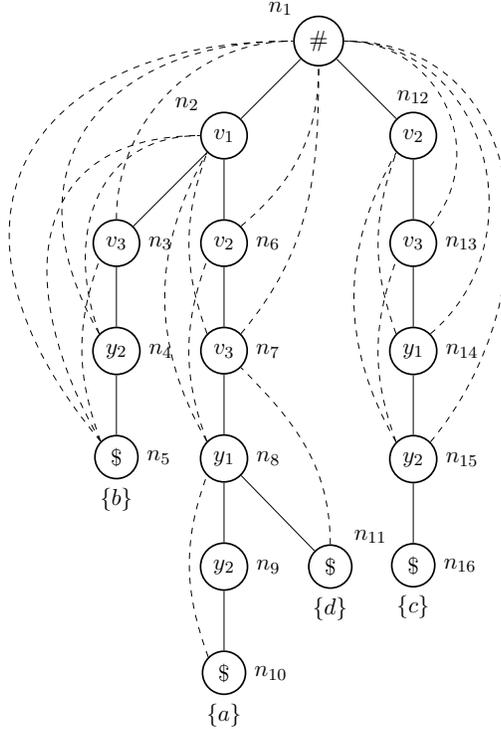
\end{examplecont}
\item{Step 9.} Convert the trie-like graph $G$ into a layered graph, call it $G'$, using either Chen's Algorithm~1 or Algorithm~3 (their improved version of Algorithm~1). A layered graph consists of nodes from the trie-like graph $G$, possibly duplicates, arranged in ``stacked layers,'' with edges from $G$ (again, possibly duplicates) going between the layers.
\begin{examplecont}
    See Figure~\ref{fig:example-layered-graph} for the layered graph $G'$ produced by Chen's Algorithm~1 applied to $G$.
    \begin{figure}
        \centering
        \begin{adjustbox}{width=\textwidth}
        \begin{tikzpicture}
            [rnode/.style={circle, draw=black, thick, minimum size=7mm},]

            \node[rnode, fill=lightgray, very thick] (c1) [label=below:{$\{a\}$}, label=right:{$n_{10}$}] {\$};
            \node[rnode] (c2) [right=of c1, label=below:{$\{b\}$}, label=right:{$n_{5}$}] {\$};
            \node[rnode, fill=lightgray, very thick] (c3) [right=of c2, label=below:{$\{c\}$}, label=right:{$n_{16}$}] {\$};
            \node[rnode] (c4) [right=of c3, label=below:{$\{d\}$}, label=right:{$n_{11}$}] {\$};

            \node[rnode, draw=none] (d1) [left=of c1] {};
            \node[rnode, draw=none] (d0) [left=of d1] {};
            \node[rnode, draw=none] (d2) [left=of d0, label=left:{Layer 1:}] {};
            \node[rnode, draw=none] (d3) [right=of c4] {};
            \node[rnode, draw=none] (d4) [right=of d3] {};
            \node[rnode, draw=none] (d5) [right=of d4] {};
            
            \node[rnode] (c5) [above=of d1, label=right:{$n_{8}$}] {$y_1$};
            \node[rnode, fill=lightgray, very thick] (c6) [above=of c1, label=right:{$n_{9}$}] {$y_2$};
            \node[rnode] (c7) [above=of c2, label=right:{$n_{4}$}] {$y_2$};
            \node[rnode, fill=lightgray, very thick] (c8) [above=of c3, label=right:{$n_{15}$}] {$y_2$};
            \node[rnode] (c9) [above=of c4, label=right:{$n_{2}$}] {$v_1$};
            \node[rnode] (c10) [above=of d3, label=right:{$n_{3}$}] {$v_3$};
            \node[rnode] (c11) [above=of d4, label=right:{$n_{7}$}] {$v_3$};
            \node[rnode] (c12) [above=of d5, label=right:{$n_{1}$}] {\#};

            \node[rnode, draw=none] (d00) [left=of c5] {};
            \node[rnode, draw=none, anchor=base] (d6) at (d2 |- c5.base) [label=left:{Layer 2:}] {};
            
            \node[rnode, fill=lightgray, very thick] (c14) [above=of c5, label=right:{$n_{14}$}] {$y_1$};
            \node[rnode, fill=lightgray, very thick] (c13) [above=of d00, label=right:{$n_{8}$}] {$y_1$};
            \node[rnode] (c19) [above=of c6, label=right:{$n_{3}$}] {$v_3$};
            \node[rnode] (c20) [above=of c7, label=right:{$n_{13}$}] {$v_3$};
            \node[rnode] (c16) [above=of c8, label=right:{$n_{2}$}] {$v_1$};
            \node[rnode] (c17) [above=of c9, label=right:{$n_{12}$}] {$v_2$};
            \node[rnode] (c21) [above=of c10, label=right:{$n_{1}$}] {\#};
            \node[rnode] (c15) [above=of c11, label=right:{$n_{2}$}] {$v_1$};
            \node[rnode] (c18) [above=of c12, label=right:{$n_{6}$}] {$v_2$};
            \node[rnode] (c22) [right=of c18, label=right:{$n_{1}$}] {\#};

            \node[rnode] (c23) [above=of c13, label=right:{$n_{2}$}] {$v_1$};
            \node[rnode, fill=lightgray, very thick] (c24) [above=of c14, label=right:{$n_{6}$}] {$v_2$};
            \node[rnode, fill=lightgray, very thick] (c25) [above=of c19, label=right:{$n_{12}$}] {$v_2$};
            \node[rnode] (c26) [above=of c20, label=right:{$n_{7}$}] {$v_3$};
            \node[rnode] (c27) [above=of c16, label=right:{$n_{13}$}] {$v_3$};
            \node[rnode] (c28) [above=of c17, label=right:{$n_{1}$}] {\#};
            \node[rnode] (c29) [above=of c21, label=right:{$n_{2}$}] {$v_1$};
            \node[rnode] (c30) [above=of c15, label=right:{$n_{12}$}] {$v_2$};
            \node[rnode] (c31) [above=of c18, label=right:{$n_{1}$}] {\#};

            \node[rnode] (c32) [above=of c24, label=right:{$n_{2}$}] {$v_1$};
            \node[rnode, fill=lightgray, very thick] (c33) [above=of c25, label=right:{$n_{1}$}] {\#};
            \node[rnode] (c34) [above=of c26, label=right:{$n_{6}$}] {$v_2$};
            \node[rnode] (c35) [above=of c27, label=right:{$n_{12}$}] {$v_2$};
            \node[rnode] (c36) [above=of c28, label=right:{$n_{2}$}] {$v_1$};
            \node[rnode] (c37) [above=of c29, label=right:{$n_{1}$}] {\#};

            \node[rnode] (c38) [above=of c34, label=right:{$n_{2}$}] {$v_1$};
            \node[rnode] (c39) [above=of c35, label=right:{$n_{1}$}] {\#};

            \node[rnode, draw=none, anchor=base] (d7) at (d2 |- c13.base) [label=left:{Layer 3:}] {};
            \node[rnode, draw=none, anchor=base] (d8) at (d2 |- c23.base) [label=left:{Layer 4:}] {};
            \node[rnode, draw=none, anchor=base] (d9) at (d2 |- c32.base) [label=left:{Layer 5:}] {};
            \node[rnode, draw=none, anchor=base] (d11) at (d2 |- c38.base) [label=left:{Layer 6:}] {};
            
            \draw[-] (c1.north west) -- (c5.south east);
            \draw[-, very thick] (c1.north) -- (c6.south);
            \draw[-] (c2.north) -- (c7.south);
            \draw[-] (c2.north east) -- (c9.south west);
            \draw[-] (c2.north east) -- (c10.south west);
            \draw[-] (c2.north east) -- (c12.south west);
            \draw[-, very thick] (c3.north) -- (c8.south);
            \draw[-] (c4.north west) -- (c5.south east);
            \draw[-] (c4.north east) -- (c11.south west);

            \draw[-, very thick] (c6.north west) -- (c13.south east);
            \draw[-] (c7.north east) -- (c16.south west);
            \draw[-] (c7.north west) -- (c19.south east);
            \draw[-] (c7.north east) -- (c21.south west);
            \draw[-, very thick] (c8.north west) -- (c14.south east);
            \draw[-] (c8.north east) -- (c17.south west);
            \draw[-] (c8.north west) -- (c20.south east);
            \draw[-] (c8.north east) -- (c21.south west);
            \draw[-] (c10.north east) -- (c15.south west);
            \draw[-] (c10.north east) -- (c22.south west);
            \draw[-] (c11.north) -- (c15.south);
            \draw[-] (c11.north east) -- (c18.south west);
            \draw[-] (c11.north east) -- (c22.south west);

            \draw[-] (c13.north) -- (c23.south);
            \draw[-, very thick] (c13.north east) -- (c24.south west);
            \draw[-] (c13.north east) -- (c26.south west);
            \draw[-, very thick] (c14.north east) -- (c25.south west);
            \draw[-] (c14.north east) -- (c27.south west);
            \draw[-] (c14.north east) -- (c28.south west);
            \draw[-] (c19.north east) -- (c29.south west);
            \draw[-] (c19.north east) -- (c31.south west);
            \draw[-] (c20.north east) -- (c30.south west);
            \draw[-] (c20.north east) -- (c31.south west);

            \draw[-] (c24.north) -- (c32.south);
            \draw[-, very thick] (c24.north east) -- (c33.south west);
            \draw[-, very thick] (c25.north) -- (c33.south);
            \draw[-] (c26.north) -- (c34.south);
            \draw[-] (c26.north east) -- (c36.south west);
            \draw[-] (c26.north east) -- (c37.south west);
            \draw[-] (c27.north) -- (c35.south);
            \draw[-] (c27.north east) -- (c37.south west);

            \draw[-] (c34.north) -- (c38.south);
            \draw[-] (c34.north east) -- (c39.south west);
            \draw[-] (c35.north) -- (c39.south);
      
            \node[draw, dashed, inner sep=1mm, fit=(c6) (c7) (c8)] {};
            \node[draw, dashed, inner sep=1mm, fit=(c10) (c11)] {};
            \node[draw, dashed, inner sep=1mm, fit=(c13) (c14)] {};
            \node[draw, dashed, inner sep=1mm, fit=(c19) (c20)] {};
            \node[draw, dashed, inner sep=1mm, fit=(c24) (c25)] {};
            \node[draw, dashed, inner sep=1mm, fit=(c26) (c27)] {};
            \node[draw, dashed, inner sep=1mm, fit=(c34) (c35)] {};
        \end{tikzpicture}
        \end{adjustbox}
        \caption{The layered graph for the running example. The shaded nodes and bolded edges form a rooted subgraph, as detailed later in Section~\ref{s:search-find-subset}.} \label{fig:example-layered-graph}
    \end{figure}
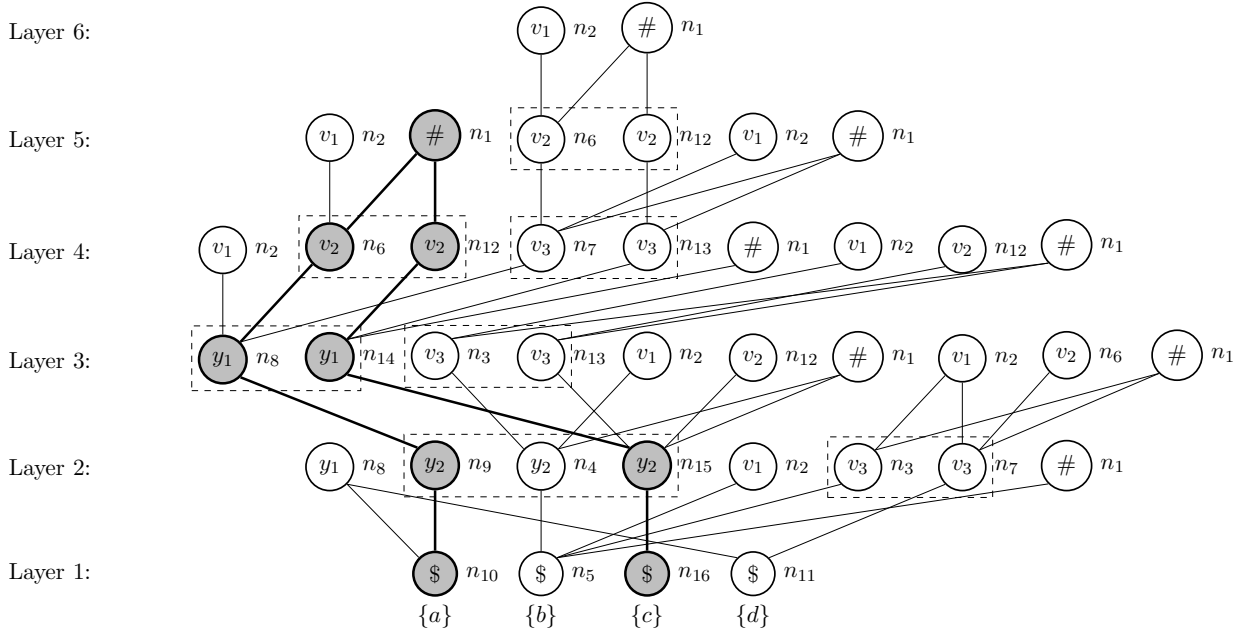
\end{examplecont}
\item{Step 10.} Apply Chen's Algorithm~2 to $G'$, which Chen claims gives in its output the maximum number of satisfied clauses in the original 2-CNF formula~$F$~\cite{chen:2-maxsat-solved-in-p-time}.
\begin{examplecont}
    Chen's Algorithm~2 examines $G'$ and returns $2$ as the maximum number of satisfied clauses in $F$. 
\end{examplecont}
\end{description}

\section{Analysis of Chen's Algorithm~1 (\textit{SEARCH}) and Algorithm~2 (\textit{findSubset})}
\label{s:search-find-subset}

Chen's Algorithm~1 takes a trie-like graph $G$ as input, uses $G$ to form a layered graph called $G'$, and then returns Algorithm~2 applied to $G'$, which itself returns a tuple of values that includes the maximum number of satisfied clauses~\cite{chen:2-maxsat-solved-in-p-time}. We detail here, informally, how Chen's Algorithm~1 and Algorithm~2 work, with changes to notation for clarity.

Algorithm~1 first initializes the layered graph $G'$ to be the set of all leaf nodes in the trie-like graph $G$, and subsequently pushes this set onto an empty stack $S$. In the main loop of the algorithm, a set $g$ is popped off of $S$. For each node $n$ in $g$, every parent node $p$ of $n$ and the corresponding edge from $n$ to $p$ are added to $G'$. All of the parents of nodes in $g$ are then grouped into sets by their variable label,\footnote{We note here that nodes in a layer form a group only if the nodes they are parents of exist in the same group in the next lowest layer. For example, looking at the running example layered graph of Figure~\ref{fig:example-layered-graph}, in Layer 4 the shaded nodes $n_6$ and $n_{12}$ (both labeled $v_2$) form a group as they are parents of nodes that form a group in Layer 3 (which are also shaded). However, another instance of the node $n_{12}$ that is in Layer 4 (on the right side of the layer) cannot group with the shaded $n_6$ and $n_{12}$ as it is not a parent of the shaded group in Layer 3. As a consequence of this behavior, layers in $G'$ can contain multiple instances of the same node (as seen in the case of $n_{12}$ in Layer 4).} and sets with at least two nodes are pushed onto $S$. This loop repeats until $S$ is empty. Finally, Algorithm~1 returns Algorithm~2 applied to the completed $G'$. Chen purports that Algorithm~1 correctly creates a $G'$ of ``rooted subgraphs''~\cite{chen:2-maxsat-solved-in-p-time}.

Algorithm~2 works by examining $G'$ to find rooted subgraphs. A rooted subgraph consists of a root node (a node without parents), internal nodes, and one or more leaf nodes that each correspond to one or more 2-DNF conjunctions, with the variable labels of the nodes along a path from the root to a leaf forming a satisfying assignment to the conjunction(s) at the leaf. An example of a rooted subgraph can be seen in the layered graph depicted in Figure~\ref{fig:example-layered-graph}. In Algorithm~2, the rooted subgraphs are derived from $G'$, and for each rooted subgraph the subset of satisfied 2-DNF conjunction is found. Once finished, the algorithm outputs a tuple that includes the maximum number of 2-DNF conjunctions satisfied by a truth assignment within a rooted subgraph, which Chen claims is equal to the maximum number of satisfied clauses in the original 2-CNF formula~\cite{chen:2-maxsat-solved-in-p-time}.

\subsection{Chen's Algorithm~1}
\label{s:chens-algo-one}
In this subsection, we give three counterexample 2-CNF formulas that result in Chen's Algorithm~1 reporting incorrect results, which demonstrate that Algorithm~1 is flawed. The first and second of these counterexamples can be extended to infinitely many counterexamples. In these counterexamples, we assume that Chen's Algorithm~2 is correctly implemented, i.e.\ that it behaves as described by Chen.

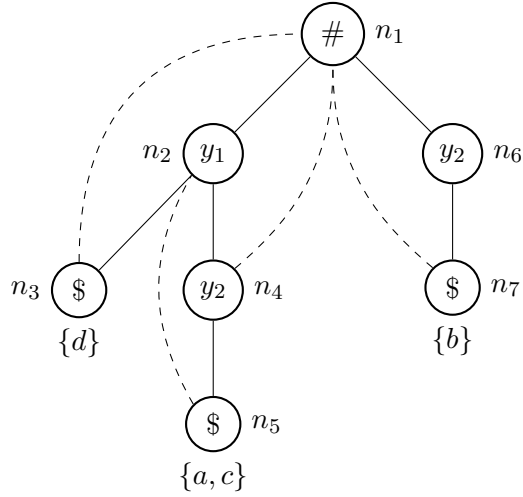
\begin{figure}[H]
    \centering
    \begin{tikzpicture}
        [rnode/.style={circle, draw=black, thick, minimum size=7mm},]

        \node[rnode] (c1) [label=right:{$n_1$}] {\#};
        \node[rnode] (c2) [below left=of c1, label=left:{$n_2$}] {$y_1$};
        \node[rnode] (c4) [below=of c2, label=right:{$n_4$}] {$y_2$};
        \node[rnode] (c5) [below=of c4, label=below:{$\{a,c\}$}, label=right:{$n_5$}] {\$};
        \node[rnode] (c3) [left=of c4, label=below:{$\{d\}$}, label=left:{$n_3$}] {\$};
        \node[rnode] (c6) [below right=of c1, label=right:{$n_6$}] {$y_2$};
        \node[rnode] (c7) [below=of c6, label=below:{$\{b\}$}, label=right:{$n_7$}] {\$};
        
        \draw[-] (c1.south west) -- (c2.north east);
        \draw[-] (c2.south west) -- (c3.north east);
        \draw[-] (c2.south) -- (c4.north);
        \draw[-] (c4.south) -- (c5.north);
        \draw[-] (c1.south east) -- (c6.north west);
        \draw[-] (c6.south) -- (c7.north);
        
        \draw[dashed, looseness=1.2] (c3.north) to[out=90,in=180] (c1.west);
        \draw[dashed] (c5.north west) to[out=120,in=240] (c2.south west);
        \draw[dashed] (c4.north east) to[out=40,in=-90] (c1.south);
        \draw[dashed] (c7.north west) to[out=140,in=-90] (c1.south);
    \end{tikzpicture}
    \caption{The trie-like graph for Counterexample 1.} \label{fig:c1-a1-trie}
\end{figure}

\begin{figure}[H]
    \centering
    \begin{tikzpicture}
        [rnode/.style={circle, draw=black, thick, minimum size=7mm},]

        \node[rnode, fill=lightgray, very thick] (c1) [label=below:{$\{d\}$}, label=right:{$n_{3}$}] {\$};
        \node[rnode, fill=lightgray, very thick] (c2) [right=of c1, label=below:{$\{a,c\}$}, label=right:{$n_{5}$}] {\$};
        \node[rnode] (c3) [right=of c2, label=below:{$\{b\}$}, label=right:{$n_{7}$}] {\$};
        \node[rnode] (c4) [above=of c1, label=right:{$n_{1}$}] {\#};
        \node[rnode, fill=lightgray, very thick] (c5) [right=of c4, above=of c2, label=right:{$n_{2}$}] {$y_1$};
        \node[rnode] (c6) [right=of c5, above=of c3, label=right:{$n_{4}$}] {$y_2$};
        \node[rnode] (c7) [right=of c6, label=right:{$n_{6}$}] {$y_2$};
        \node[rnode] (c8) [above=of c6, label=right:{$n_{1}$}] {\#};
        \node[rnode] (c9) [right=of c8, label=right:{$n_{2}$}] {$y_1$};

        \node[rnode, draw=none] (d1) [left=of c1, label=left:{Layer 1:}] {};
        \node[rnode, draw=none, anchor=base] (d2) at (d1 |- c4.base) [label=left:{Layer 2:}] {};
        \node[rnode, draw=none, anchor=base] (d4) at (d1 |- c8.base) [label=left:{Layer 3:}] {};

        \draw[-] (c1.north) -- (c4.south);
        \draw[-, very thick] (c1.north east) -- (c5.south west);
        \draw[-, very thick] (c2.north) -- (c5.south);
        \draw[-] (c2.north east) -- (c6.south west);
        \draw[-] (c3.north east) -- (c7.south west);
        \draw[-] (c3.north west) -- (c4.south east);
        \draw[-] (c6.north) -- (c8.south);
        \draw[-] (c6.north east) -- (c9.south west);
        \draw[-] (c7.north west) -- (c8.south east);
        
        \node[draw, dashed, inner sep=2mm, fit=(c6) (c7)] {};
    \end{tikzpicture}
    \caption{The layered graph for Counterexample~1, as generated by Chen's Algorithm~1.} \label{fig:c1-a1-layered-graph}
\end{figure}
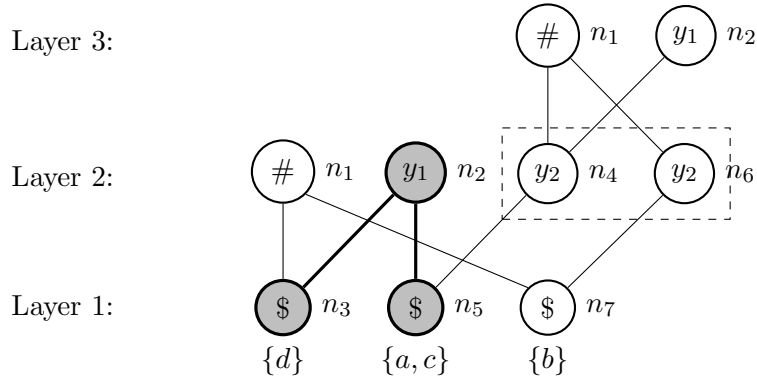

For the first counterexample (Counterexample~1), consider the 2-CNF formula
\begin{equation*}
    (\neg v_1 \vee \neg v_1) \wedge (\neg v_1 \vee \neg v_1).
\end{equation*}
This formula clearly can have at most 2 satisfied clauses. The first step is to convert the 2-CNF formula into a new 2-DNF formula as described by Chen, with the added variables being of the form $y_i$:
\begin{equation*}
    \underbrace{(\neg v_1 \wedge y_1)}_ \text{Conjunction a} \vee \underbrace{(\neg v_1 \wedge \neg y_1)}_ \text{Conjunction b} \vee \underbrace{(\neg v_1 \wedge y_2)}_ \text{Conjunction c} \vee \underbrace{(\neg v_1 \wedge \neg y_2)}_ \text{Conjunction d}.
\end{equation*}
Next, we convert the 2-DNF formula into sequences of variables, while removing the negated variables (including all occurrences of $v_1$), and sorting each sequence by a global variable ordering. This global ordering is determined by the frequencies with which the variables appear across all sequences, with the most frequent variable (the variable appearing in the most sequences) coming first in the ordering (Chen claims that any global ordering can be used~\cite{chen:2-maxsat-solved-in-p-time}, but that one based on the frequencies of the variables improves the efficiency of Algorithm~1; thus we also order variables based on their frequencies to match the process used in Chen's paper). Crucially, Chen specifies that ties in the global ordering are broken arbitrarily~\cite{chen:2-maxsat-solved-in-p-time}. All of the variables that appear in our sequences occur with the same frequency, so we can arrange a variable ordering that exploits flaws in Chen's Algorithm~1. Specifically, we use the global ordering $y_1 > y_2 > v_1$ ($v_1$ has a frequency of 0), where $k > j$ indicates that variable $k$ comes before (to the left of) variable $j$ in the sequence, to create the following sorted sequences (listed in the same order as the corresponding conjunctions in the 2-DNF formula)
\begin{align}
    &\#.y_1.(y_2,*).\$ \tag{a}
    \\
    &\#.(y_2,*).\$ \tag{b}
    \\
    &\#.(y_1,*).y_2.\$ \tag{c}
    \\
    &\#.(y_1,*).\$ \tag{d}
\end{align}
Note that sequences $a$ and $c$ contain exactly the same variables, and that the variables of sequences $b$ and $d$ can all be spanned. From each sorted sequence we form a $p$-graph and $p$*-graph, which are then merged together to form the trie-like graph of Figure~\ref{fig:c1-a1-trie}.
We can then run Algorithm~1 on this trie-like graph to create the layered graph shown in Figure~\ref{fig:c1-a1-layered-graph} (the dashed boxes indicate a grouping of two or more nodes labeled with the same variable).
The layered graph clearly contains a rooted subgraph (shaded) that has three satisfied conjunctions among its roots (in fact, the layered graph contains two such rooted subgraphs). Therefore, Algorithm~2 will report the maximum number of satisfied clauses to be three, which is incorrect.

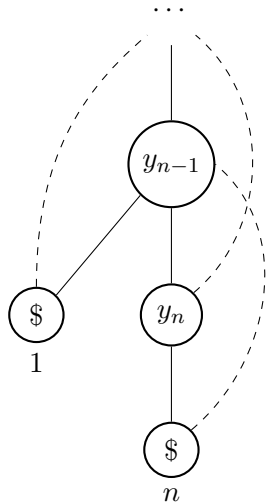
\begin{figure}[H]
    \centering
    \begin{tikzpicture}
        [rnode/.style={circle, draw=black, thick, minimum size=7mm},]

        \node[rnode, draw=none] (c1) {$\cdots$};
        \node[rnode] (c2) [below=of c1] {$y_{n-1}$};
        \node[rnode] (c3) [below=of c2] {$y_n$};
        \node[rnode] (c4) [below=of c3, label=below:{$n$}] {\$};
        \node[rnode] (c5) [left=of c3, label=below:{1}] {\$};
        
        \draw[-] (c1.south) -- (c2.north);
        \draw[-] (c2.south) -- (c3.north);
        \draw[-] (c3.south) -- (c4.north);
        \draw[-] (c2.south west) -- (c5.north east);
        \draw[dashed] (c4.north east) to[out=40,in=-40] (c2.east);
        \draw[dashed] (c5.north) to[out=90,in=220] (c1.south west);
        \draw[dashed] (c3.north east) to[out=40,in=-40] (c1.south east);
    \end{tikzpicture}
    \caption{A section of a trie-like graph.} \label{fig:partial-trie}
\end{figure}

This counterexample can be extended to infinite counterexamples of the form
\begin{equation*}
    \underbrace{(\neg v_1 \vee \neg v_1) \wedge (\neg v_1 \vee \neg v_1) \wedge \cdots \wedge (\neg v_1 \vee \neg v_1)}_ \text{Total of $n$ clauses},
\end{equation*}
where $n>1$. For a boolean formula of the above form with $n$ clauses assume without loss of generality that the 2-DNF conjunctions were created in the same order as their corresponding 2-CNF clauses, and that the variables in the 2-DNF conjunctions that were added during the 2-CNF to 2-DNF conversion are numbered in order they were added. Additionally, assume that the order of the sequences matches that of the 2-DNF conjunctions. Then if the sequences are sorted using a global ordering that has the additional variables in numerical order and $v_1$ is last ($y_1 > y_2 > \cdots > y_n > v_1$), half of the sequences (every odd one, for a total of $n$), call them the ``full'' sequences, will contain every variable except $v_1$. One sequence (the final sequence), call it the ``prefix'' sequence, will contain every variable except $v_1$ and $y_n$, the numerically last variable added during the 2-CNF to 2-DNF conversion. When creating the trie-like graph, the full sequences and the prefix sequences will share a trie branch as depicted in Figure~\ref{fig:partial-trie}, with the number below a leaf indicating the number of conjunctions associated with that leaf.
Chen's search algorithm will create a layered graph that contains a rooted subgraph with three nodes: two leaf nodes with the conjunctions of the full and prefix sequences among them, and a root node labeled with the variable $y_{n-1}$. Clearly, Algorithm~2 will report the maximum number of satisfied clauses to be $n+1$, despite the 2-CNF formula containing only $n$ clauses. There may exist multiple such ``illegal'' rooted subgraphs in a layered graph.

We briefly mention here our second counterexample (Counterexample~2), the \mbox{2-CNF} formula $(v_1 \vee v_1) \wedge (v_1 \vee v_1)$, which is a slight variation of that of Counterexample~1.

The steps of this counterexample are essentially the same as for Counterexample~1, except that the global ordering must have $v_1$ at the front rather than the end, and so $v_1$ will appear at the front of all of the sequences. The trie-like graph will contain the same section as depicted in Figure~\ref{fig:partial-trie}, and so Algorithm~1 will exhibit an identical issue as to Counterexample~1, incorrectly reporting the maximum number of satisfied clauses to be three. We leave it to the reader to see that, as with Counterexample~1, this can also be extended to infinitely many counterexamples.

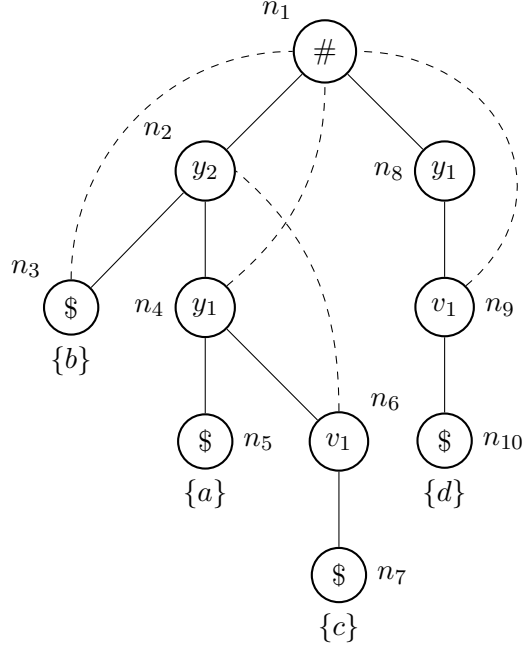
\begin{figure}
    \centering
    \begin{tikzpicture}
        [rnode/.style={circle, draw=black, thick, minimum size=7mm},]

        \node[rnode] (c1) [label=above left:{$n_{1}$}] {\#};
        \node[rnode] (c2) [below left=of c1, label=above left:{$n_{2}$}] {$y_2$};
        \node[rnode] (c4) [below=of c2, label=left:{$n_{4}$}] {$y_1$};
        \node[rnode] (c3) [left=of c4, label=below:{$\{b\}$}, label=above left:{$n_{3}$}] {\$};
        \node[rnode] (c5) [below=of c4, label=below:{$\{a\}$}, label=right:{$n_{5}$}] {\$};
        \node[rnode] (c6) [right=of c5, label=above right:{$n_{6}$}] {$v_1$};
        \node[rnode] (c7) [below=of c6, label=below:{$\{c\}$}, label=right:{$n_{7}$}] {\$};
        \node[rnode] (c8) [below right=of c1, label=left:{$n_{8}$}] {$y_1$};
        \node[rnode] (c9) [below=of c8, label=right:{$n_{9}$}] {$v_1$};
        \node[rnode] (c10) [below=of c9, label=below:{$\{d\}$}, label=right:{$n_{10}$}] {\$};
        
        \draw[-] (c1.south west) -- (c2.north east);
        \draw[-] (c2.south west) -- (c3.north east);
        \draw[-] (c2.south) -- (c4.north);
        \draw[-] (c4.south) -- (c5.north);
        \draw[-] (c4.south east) -- (c6.north west);
        \draw[-] (c6.south) -- (c7.north);
        \draw[-] (c1.south east) -- (c8.north west);
        \draw[-] (c8.south) -- (c9.north);
        \draw[-] (c9.south) -- (c10.north);
        
        \draw[dashed] (c3.north) to[out=90,in=180] (c1.west);
        \draw[dashed] (c4.north east) to[out=40,in=-90] (c1.south);
        \draw[dashed] (c6.north) to[out=90,in=-40] (c2.east);
        \draw[dashed, looseness=1.4] (c9.north east) to[out=40,in=0] (c1.east);
    \end{tikzpicture}
    \caption{The trie-like graph for Counterexample 3.} \label{fig:c3-a1-trie}
\end{figure}

\begin{figure}
    \centering
    \begin{tikzpicture}
        [rnode/.style={circle, draw=black, thick, minimum size=7mm},]

        \node[rnode] (c1) [label=below:{$\{b\}$}, label=right:{$n_{3}$}] {\$};
        \node[rnode, draw=none] (d1) [right=of c1] {};
        \node[rnode] (c2) [right=of d1, label=below:{$\{a\}$}, label=right:{$n_{5}$}] {\$};
        \node[rnode] (c3) [right=of c2, label=below:{$\{c\}$}, fill=lightgray, very thick, label=right:{$n_{7}$}] {\$};
        \node[rnode] (c4) [right=of c3, label=below:{$\{d\}$}, fill=lightgray, very thick, label=right:{$n_{10}$}] {\$};
        \node[rnode] (c5) [above=of c1, label=right:{$n_{2}$}] {$y_2$};
        \node[rnode] (c6) [above=of d1, label=right:{$n_{1}$}] {\#};
        \node[rnode] (c7) [above=of c2, label=right:{$n_{4}$}] {$y_1$};
        \node[rnode] (c8) [above=of c3, fill=lightgray, very thick, label=right:{$n_{6}$}] {$v_1$};
        \node[rnode] (c9) [above=of c4, fill=lightgray, very thick, label=right:{$n_{9}$}] {$v_1$};
        \node[rnode] (c10) [above=of c8, fill=lightgray, very thick, label=right:{$n_{4}$}] {$y_1$};
        \node[rnode] (c11) [above=of c9, fill=lightgray, very thick, label=right:{$n_{8}$}] {$y_1$};
        \node[rnode] (c12) [above=of c7, label=right:{$n_{2}$}] {$y_2$};
        \node[rnode] (c13) [right=of c11, label=right:{$n_{1}$}] {\#};
        \node[rnode] (c14) [above=of c11, fill=lightgray, very thick, label=right:{$n_{1}$}] {\#};
        \node[rnode] (c15) [above=of c10, label=right:{$n_{2}$}] {$y_2$};

        \node[rnode, draw=none] (d2) [left=of c1, label=left:{Layer 1:}] {};
        \node[rnode, draw=none] (d3) [left=of c5, label=left:{Layer 2:}] {};
        \node[rnode, draw=none] (d4) [above=of d3, label=left:{Layer 3:}] {};
        \node[rnode, draw=none] (d5) [above=of d4, label=left:{Layer 4:}] {};
    
        \draw[-] (c1.north) -- (c5.south);
        \draw[-] (c1.north east) -- (c6.south west);
        \draw[-] (c2.north) -- (c7.south);
        \draw[-, very thick] (c3.north) -- (c8.south);
        \draw[-, very thick] (c4.north) -- (c9.south);
        \draw[-] (c8.north west) -- (c12.south east);
        \draw[-, very thick] (c8.north) -- (c10.south);
        \draw[-, very thick] (c9.north) -- (c11.south);
        \draw[-] (c9.north east) -- (c13.south west);
        \draw[-] (c10.north) -- (c15.south);
        \draw[-, very thick] (c10.north east) -- (c14.south west);
        \draw[-, very thick] (c11.north) -- (c14.south);

        \node[draw, dashed, inner sep=2mm, fit=(c8) (c9)] {};
        \node[draw, dashed, inner sep=2mm, fit=(c10) (c11)] {};
    \end{tikzpicture}
    \caption{The layered graph for Counterexample 3, as generated by Chen's Algorithm~1.} \label{fig:c3-a1-layered-graph}
\end{figure}
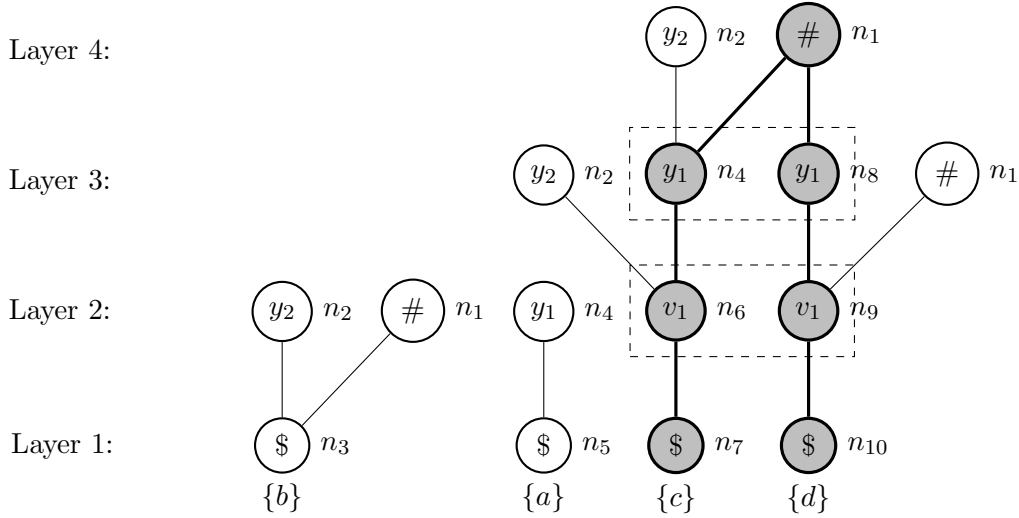

For the third counterexample (Counterexample~3), consider the following 2-CNF formula
\begin{equation*}
    (\neg v_1 \vee \neg v_1) \wedge (v_1 \vee v_1).
\end{equation*}
Unlike the previous two counterexamples, this formula is clearly not satisfiable. As with Counterexample~1, the first step is to convert the 2-CNF formula into the following 2-DNF formula:
\begin{equation*}
    \underbrace{(\neg v_1 \wedge y_1)}_ \text{Conjunction a} \vee \underbrace{(\neg v_1 \wedge \neg y_1)}_ \text{Conjunction b} \vee \underbrace{(v_1 \wedge y_2)}_ \text{Conjunction c} \vee \underbrace{(v_1 \wedge \neg y_2)}_ \text{Conjunction d}.
\end{equation*}
Next, we convert the 2-DNF formula into sequences of variables. We again ``rig'' the global ordering, exploiting the fact that ties are broken arbitrarily to create the ordering $y_2 > y_1 > v_1$ ($v_1$ must come last in the ordering), which results in the following sorted sequences (in the same order as the corresponding conjunctions of the 2-DNF formula):
\begin{align}
    &\#.(y_2,*).y_1.\$ \tag{a}
    \\
    &\#.(y_2,*).\$ \tag{b}
    \\
    &\#.y_2.(y_1,*).v_1.\$ \tag{c}
    \\
    &\#.(y_1,*).v_1.\$ \tag{d}
\end{align}
From the sorted sequences, we form the trie-like graph of Figure~\ref{fig:c3-a1-trie}.
We run Algorithm~1 on the trie-like graph, creating the layered graph of Figure~\ref{fig:c3-a1-layered-graph}.
The layered graph clearly contains a rooted subgraph (shaded) with two satisfied conjunctions among its roots. Therefore, Algorithm~2 will report the maximum number of satisfied clauses to be two, which is incorrect.

One issue made apparent by these counterexamples is that the spans (and the fact that variable sequences can share branches in the trie-like graph) allow Algorithm~1 to ``skip over'' variables that are ``important'' to the satisfiability of a conjunction. For instance, in the layered graph of Counterexample~1, conjunctions $a$ and $c$ are shown to be satisfied by setting $y_1$ to true and $v_1$ and $y_2$ to false, even though $y_2$ must be set to true for conjunction $c$ to be satisfied. When Algorithm~1 is searching up the trie-like graph having started at the leaf for conjunction $c$, the algorithm ``thinks'' it can skip over and ignore $y_2$ because there is a span over it, even though that span only applies to conjunction $a$, not conjunction $c$. An identical issue is present in Counterexample~3, again for conjunctions $a$ and $c$.

The three counterexamples presented here demonstrate critical flaws in Chen's Algorithm~1 and show that it does not function as intended. Given that Chen's algorithm fails on relatively simple 2-CNF formulas, we think it is at least potentially possible that more complex counterexamples might reveal additional flaws.

Interestingly, Chen does not provide any time analysis for Algorithm~1, and neither do we as we have already shown the algorithm is flawed. Later in Chen's paper, Algorithm~1 is replaced with with Algorithm~3, which they claim is an improved search algorithm~\cite{chen:2-maxsat-solved-in-p-time}. It is unclear whether Algorithm~3 is an improvement over Algorithm~1 in that it brings the runtime from exponential time to polynomial time, or whether Algorithm~1 already runs in polynomial time and Algorithm~3 simply reduces the degree of the polynomial.
As we later show in Section~\ref{s:graph-improvements}, the issue found with Algorithm~1, originating in the trie-like graph, still holds in Algorithm~3, which also takes a trie-like graph as input.

\subsection{Chen's Algorithm~2}
\label{s:chens-algo-two}

Chen's Algorithm~2 takes a layered graph $G'$ and outputs a tuple that includes the largest subset of conjunctions satisfied by a certain truth assignment. The algorithm works by determining the subset $D'$ of satisfied conjunctions in each rooted subgraph $G_v$ and updating the maximum size for each iteration.

First, it should be noted that in their paper~\cite{chen:2-maxsat-solved-in-p-time}, Chen uses $n$ to refer to the number of clauses and $m$ to refer to the number of variables in a boolean formula; however, this is only mentioned briefly in their introduction. In Section~3 of Chen's paper~\cite{chen:2-maxsat-solved-in-p-time}, the author themselves seems to confuse these two values by incorrectly claiming that the maximum number of edges in the trie-like graph is in $O(n^2)$ (but this is fixed later in Chen's Section~5). Additionally, in Chen's introduction, it is stated that $n$ and $m$ refer to the 2-CNF formula $C$, however during our analysis of their paper, we found that these values seem to refer to the 2-DNF formula $D$. Since the number of clauses in $D$ is equal to twice the number of clauses in $C$, this does not directly impact the complexity analysis in terms of Big-Oh (but should be pointed out regardless). However, it can easily be verified that $D$ will have at exactly the number of variables in $C$ plus the number of clauses in $C$, which may impact the complexity. To avoid confusion in terms of the complexity analysis, we assume $n$ and $m$ refer to the number of clauses and variables in a 2-DNF boolean formula, and we use $n_0$ and $m_0$ to refer to the number of clauses and variables in a 2-CNF boolean formula.

Initially, we want to find the upper bounds of Chen's algorithm on the size of the layered representation for any trie-graph $G$.

\begin{proposition}
For any 2-DNF boolean formula of $n$ clauses and $m$ variables that is in the format specified by Chen, its trie-like graph $G$ will have at most $n(m+2)-1$ vertices and $\frac{(m+2)(m+1)n}{2}$ edges, and its layered-graph $G'$ will have at most $(n(m+2)-1)(m+2)$ vertices and $\frac{(m+2)(m+1)^2n}{2}$ edges.
\end{proposition}
\begin{myproof}
    We will use similar notation to that of Chen's paper, and will reference steps from Section~\ref{s:overview}.

    Let $C = C_1 \wedge C_2 \wedge \cdots \wedge C_{n_0}$ be a 2-CNF formula with $n_0$ clauses and $m_0$ variables, where $m_0 \leq 2n_0$ and each $C_i$ is a clause of the form $y_i \vee y_j$. By converting it to a 2-DNF formula in Step~1, the number of clauses will be $n = 2n_0$ and the number of variables will be $m = m_0+n_0$. We then convert each conjunction into a variable sequence and pad with $(c_j,*)$, where $(c_j,*) = c_j \vee \neg c_j$ and get at most $2n_0(n_0+m_0)=mn$ variables before converting to $p$-graph, finishing Step~2. After that, by Steps 3 and 4, for each clause, we will have $2$ new symbols, $\#$ and $\$$, making at most $mn+2n = (m+2)n$ variables and $n$ clauses.

    For Step~5, we convert each of the sorted sequence into a $p$-graph. Given any clause $D_i$ with at most $m+2$ variables, its $p$-graph $P_i$ will have at most $(m+1)+m = 2m+1$ edges. Hence, considering all of the $p$-graphs, we will get $n$ different graphs, with each of them having at most $2m+1$ edges and $m+2$ vertices. In total, there will be at most $n(2m+1)$ edges and $n(m+2)$ vertices.
    
    For Step~6, we convert each $p$-graph into a $p^*$-graph. We see that the total number of vertices will also be $n(m+2)$, while for each $p$-graph, the worst case is to have any two vertices connected to each other. Thus for each $p$-graph, its corresponding $p^*$-graph will have at most $\frac{(m+2)(m+1)}{2}$ edges, where we consider that there exists an edge between any two vertices. Hence in total we have at most $\frac{(m+2)(m+1)n}{2}$ edges and $n(m+2)$ vertices.

    For Step~7, we merge the $p$-graphs into a single trie $T$. We see that, without counting the spans of the $p$-graphs, the worst case for creating $T$ would involve the $p$-graphs only sharing their initial $\#$-vertex. Then, every other edge from the $p$-graphs would need to be added to $T$, as none of these edges would be duplicates of each other. Hence, $T$ will have at most $n(m+2)-1$ vertices and $(m+1)n$ edges, since the trie will only have edges between consecutive vertices.
 
    For Step~8, we convert $T$ into a trie-like graph $G$ by adding the spans from all of the $p^*$-graphs generated in Step~6 to $T$. We see that after this process, the number of edges cannot exceed the total number of edges in the $p^*$-graphs so we will have at most $n(m+2)-1$ vertices and $\frac{(m+2)(m+1)n}{2}$ edges.

    Finally, we want to convert $G$ into the layered graph $G'$. Chen achieves this by searching $G$ from the bottom up, and connecting the nodes in each layer of $G'$ to all of its parents~\cite{chen:2-maxsat-solved-in-p-time}. We can see that the depth of the layered graph is at most $m+2$ minus the largest number of vertices on each of the $p$-graphs, which can only be achieved if we move through the main path of $G$ from the bottom of the graph to the top. Note that the condition for vertex $x$ in layer $i$ and vertex $y$ in layer $i+1$ to have an edge in $G'$ is that the group of $x$ at layer $i$ has at least two elements and there is an edge between $x$ and $y$ in $G$. Hence, between each layer, the number of edges is at most the total number of edges in $G$ (which is $\frac{(m+2)(m+1)n}{2}$), so in total there will be at most $\frac{(m+2)(m+1)^2n}{2}$ edges in $G'$. Moreover, we see that the number of vertices in each layers is also at most $n(m+2)-1$ so we have at most $(n(m+2)-1)(m+2)$ vertices in $G'$.

    In conclusion, the layered graph generated by Algorithm~1 has at most $(n(m+2)-1)(m+2)$ vertices and $\frac{(m+2)(m+1)^2n}{2}$ edges. 
\end{myproof}
From Proposition~1, we see that the layered graph will have at most a polynomial number of vertices and edges with respect to the number of clauses and variables in its corresponding 2-DNF formula. Additionally, we see that in Algorithm~2 we are iterating over each rooted subgraph $G_v$ to find a maximized number of satisfied conjunctions. Since there are only a polynomial number of iterations, if we can show that the main loop of Algorithm~2 runs in polynomial time, we will have shown that all of Algorithm~2 runs in polynomial time.

Suppose $\tau(G)$ is the amount of time Algorithm~2 takes to run on the trie-like graph $G$. Then we have
\begin{equation*}
\tau(G) = \sum_{v} (\tau'(G_v)+O(1)) = \left(\sum_{v} \tau'(G_v)\right)+O(n),
\end{equation*}
where $\tau'(G_v)$ is the time complexity to determine the set of satisfied conjunctions in $G_v$. However, Chen does not detail the runtime of this process. If Algorithm~2 tries to find an exact satisfiable set for $G_v$, then this algorithm is trying to solve the SAT problem, which is known to be \mbox{NP-complete}~\cite{gar-joh-sto:simplified-np-complete-problems}. Note that as we are converting to the layered-graph, the number of terms in each of the clauses does not have an upper bound. Hence, unless the boolean function has a ``nice'' form, we are unable to determine what the actual runtime is. On the other hand, if 
the algorithm only tries to run a depth-first search on each of the rooted nodes and calculate the size of the $\$$-leaves, then the algorithm runs in polynomial time.
Nonetheless, as we have mentioned in Section~\ref{s:chens-algo-one}, Chen's algorithm does not always return the correct answer.

\section{Analysis of Chen's Graph Improvements}
\label{s:graph-improvements}
We now detail Chen's proposed improvements to generating layered graphs as described in their technical report~\cite{chen:2-maxsat-solved-in-p-time}. Although this section and the subsequent section (Section~\ref{s:complexity}) are not applicable to Chen's conference version~\cite{chen:2-maxsat-solved-in-p-time-conf}, we have already provided examples where the algorithm detailed in that paper reports incorrect results. We will also detail further issues regarding Chen's conference version in Section~\ref{s:examples-over-proofs}

In Chen's Section~4, significant changes are made to Algorithm~1, the algorithm that generates the layered graph $G'$, to create Algorithm~3. Chen claims that their changes improve the runtime of Algorithm~1~\cite{chen:2-maxsat-solved-in-p-time}. These changes include eliminating duplicate occurrences of nodes in the layered graph $G'$, and creating new structures to search $G'$ (namely, Chen creates what they call ``reachable subsets through spans'' and ``upper boundaries'' to aid Algorithm~3)~\cite{chen:2-maxsat-solved-in-p-time}. As we will show, Chen fails to provide formal instructions on how to generate these new structures, and only gives definitions and examples of what values they should take on. 

In Chen's Subsection 4.A titled ``Redundancy analysis''~\cite{chen:2-maxsat-solved-in-p-time}, Chen gives three cases that they claim can identify whether a duplicate node will be created in a layer of $G'$ when traversing the trie-like graph $G$. These cases depend on knowing how the structure of the trie-like graph $G$ will effect the layered graph $G'$; specifically, they depend on whether the node(s) in $G$ that cause a duplicate node to appear in $G'$ are contained on one or more branches of $G$. While Chen claims these cases can be used to eliminate redundant nodes in the same layer of $G'$, this approach does not prevent multiple occurrences of the same nodes from appearing in other layers of $G'$ (which may lead to a higher runtime), and lumps special cases that don't reflect Case~1 or Case~2 into a general third case without explanation. Additionally, while merging these nodes reduces the number of nodes in each level (effectively bounding each level to have at most $O(nm)$ nodes), this approach does not reduce the number of edges in $G'$, which is potentially problematic as the recursive algorithm is based on searching subsets of $G'$.

From these cases that claim to determine when duplicate nodes appear in a layer of $G'$, Chen changes the way the layered graph $G'$ is generated in an attempt to make searching $G'$ more computationally efficient. To achieve this, a subset of nodes called a reachable subset through spans is defined for a node $u$ that lies on the main trie path between the root node and a repeated node $v$ of Chen's Case~1 (where $u \neq v$), and a clause~$c$~\cite{chen:2-maxsat-solved-in-p-time}. A reachable subset of $u$ through spans with label $c$ (which Chen denotes as $RS_{u}[c]$) is then defined to be ``[the set of] nodes with a same label $c$ in different subgraphs in $G[v]$ (subgraph rooted at $v$) and reachable from $u$ through a span''~\cite{chen:2-maxsat-solved-in-p-time}. It is required that $u$ is a node on a path in the trie-like graph between the root and a repeated node of Case~1 labeled $v$. It should be noted that Chen occasionally abuses this notation in their paper by omitting the label $c$ (as can be seen in their statement ``$RS_{v1} = \emptyset$''~\cite{chen:2-maxsat-solved-in-p-time}), which is unclear as $c$ is required to define this structure.

Chen's definition of a reachable subset through spans is also confusing for a set that is supposed to improve their algorithm's runtime, as it seems to require the $G'$ from Algorithm~1 to be (at least partially) generated as the rooted subgraphs that are defined in $G'$ need to be searched. However, while an individual rooted subgraph could be generated by traversing the spans in $G$, this would require an extra search algorithm that Chen fails to define~\cite{chen:2-maxsat-solved-in-p-time}. Another concern about the reachable subset through spans definition is that it requires a node of Case~1 to be found between two arbitrary points, as the specific node $v$ does not appear in the final structure. 

This concept of a reachable subset through spans is then extended into a different set of nodes that Chen calls upper boundaries~\cite{chen:2-maxsat-solved-in-p-time}. The definition for these upper boundaries again depends on rooted subgraphs of $G$ which are not defined in the paper. Additionally, generating even one upper bound requires both generating and searching an arbitrarily large number of reachable spans, which leads us to doubt that this method is more efficient than Chen's Algorithm~1. While it is claimed that these upper boundaries and reachable subsets through spans lead to a more efficient algorithm~\cite{chen:2-maxsat-solved-in-p-time}, no proof of this is ever provided by Chen, and the overall complexity for generating these structures to remove duplicate nodes in a single layer is never analyzed. 

\begin{figure}
    \centering
    \begin{tikzpicture}
        [rnode/.style={circle, draw=black, thick, minimum size=7mm},]

        \node[rnode] (c1) [label=below:{$\{b\}$},label=right:{$n_7$}, fill=lightgray, very thick] {\$};
        \node[rnode] (c2) [right=of c1, label=below:{$\{d\}$},label=right:{$n_3$}] {\$};
        \node[rnode] (c3) [right=of c2, label=below:{$\{a, c\}$},label=right:{$n_5$}, fill=lightgray, very thick] {\$};
    
        \node[rnode] (c4) [above=of c1] [fill=lightgray, very thick,label=right:{$n_6$}]{$y_2$};
        \node[rnode] (c5) [above=of c2,label=right:{$n_1$}] {\#};
        \node[rnode] (c6) [above=of c3,label=right:{$n_2$}] {$y_1$};
        \node[rnode] (c7) [right=of c6, label=right:{$n_4$}] [fill=lightgray, very thick]{$y_2$};

        \node[rnode] (c8) [above=of c6,label=right:{$n_1$}] [fill=lightgray, very thick]{\#};
        \node[rnode] (c9) [above=of c7, label=right:{$n_2$}] {$y_1$};

        \node[rnode] (c10) [above=of c9, label=right:{$n_1$}] {\#};

        \node[rnode, draw=none] (d1) [left=of c1, label=left:{Layer 1:}] {};
        \node[rnode, draw=none] (d2) [left=of c4, label=left:{Layer 2:}] {};
        \node[rnode, draw=none] (d3) [above=of d2, label=left:{Layer 3:}] {};
        \node[rnode, draw=none] (d4) [above=of d3, label=left:{Layer 4:}] {};

        \draw[-, very thick] (c1.north) -- (c4.south);
        \draw[-] (c1.north east) -- (c5.south west);

        \draw[-] (c2.north east) -- (c6.south west);
        \draw[-] (c2.north) -- (c5.south);

        \draw[-] (c3.north) -- (c6.south);
        \draw[-, very thick] (c3.north east) -- (c7.south west);

        \draw[-, very thick] (c4.north east) -- (c8.south west);
        \draw[-] (c6.north) -- (c8.south);
        \draw[-, very thick] (c7.north west) -- (c8.south east);

        \draw[-] (c7.north) -- (c9.south);
        \draw[-] (c9.north) -- (c10.south);
    \end{tikzpicture}
    \caption{The layered graph for Counterexample 1, as generated by Chen's Algorithm~3.} \label{fig:c3-a3-layered-graph}
\end{figure}
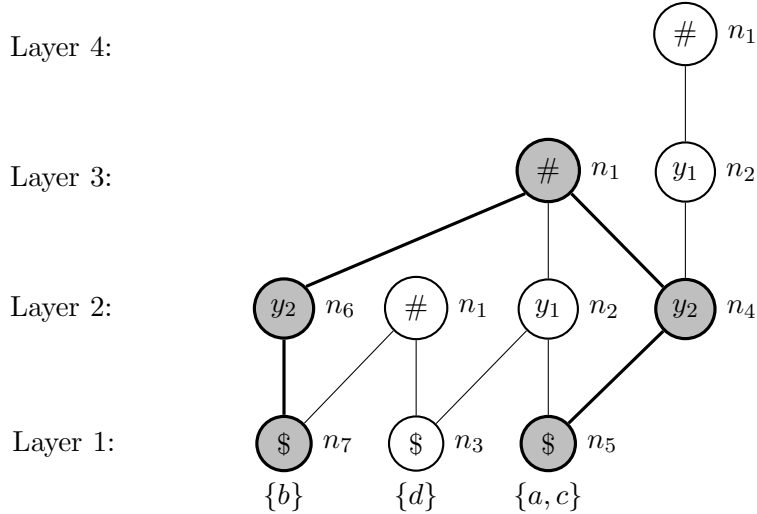

Additionally, Chen's Algorithm~3 can be shown to fail on certain inputs. For example, consider the trie-like graph we used in our Counterexample~1, depicted in Figure~\ref{fig:c1-a1-trie}. If Algorithm~3 takes this graph as input, on the first iteration of its main loop it will generate the parent node $n_1$ from $n_3$, and the parent node $n_1$ from $n_7$. Clearly this is an example of a Case~1 repeated node as described by Chen~\cite{chen:2-maxsat-solved-in-p-time}, because $n_3$ and $n_7$ appear on different main branches of $G$. Next, according to Algorithm~3 we need to generate the reachable subsets of spans relative to $n_1$. But by their definition, a reachable subset can only be defined for nodes between the root and the node before $n_1$ on the main path of $G$. Since $n_1$ is the root of $G$, such a node does not exist and we cannot calculate a reachable subset of spans (by extension, we also cannot define an upper boundary). Therefore, the recursive call will only be over the single node $n_1$, which may itself be undefined as it is unclear if all of the nodes in $G$ contain labels indicating which clauses they are a part of, or if this only applies to the leaf nodes of $G$. Similarly, the parent $n_2$ will also be generated from both $n_3$ and $n_5$. While it is not clear which duplicate node case this falls under, the reachable subset through spans will similarly be undefined as $n_1$ is the only node that fits the description of $u$, but it has no subset with more than two nodes that is relevant for generating an upper bound. Following this pattern, in all subsequently generated levels when a duplicate node appears the reachable subset through spans and upper bounds cannot be properly defined, so while duplicate nodes are still merged, the recursive calls will only be called on a single node ($n_1$). Finally, after the layered graph $G'$ is generated by Algorithm~3, it is passed to Algorithm~2 like before. As can be seen in Figure~\ref{fig:c3-a3-layered-graph}, there is still clearly a rooted subgraph that reports that three clauses can be satisfied. Therefore, despite the claimed improvements of Algorithm~3, there still exist cases where it gives an incorrect solution to the 2-MAXSAT problem.

\section{Analysis of Chen's Complexity Analysis}
\label{s:complexity}
We now remark on Chen's complexity analysis of the improved algorithm described in their technical report~\cite{chen:2-maxsat-solved-in-p-time}, which omits any complexity analysis of the unimproved algorithm. The conference version of Chen's paper~\cite{chen:2-maxsat-solved-in-p-time-conf} does include complexity analysis of the unimproved algorithm. However, its absence from the technical report and the fact that the polynomial bound of $O(n^2m^3)$ proposed in the conference version~\cite{chen:2-maxsat-solved-in-p-time-conf} is different than the bound of $O(n^3m^3)$ claimed in their updated technical report~\cite{chen:2-maxsat-solved-in-p-time} gives us reason to doubt the validity of the complexity analysis detailed in Chen's conference paper.  

Despite being a paper focused entirely on constructing an algorithm that is claimed to run in polynomial time, Chen says little on algorithmic complexity in the sections where they introduce and define their algorithms. Instead, in Chen's Section~5~\cite{chen:2-maxsat-solved-in-p-time}, the complexity of both the proposed construction of the trie-like graph $G$ and Algorithm~3 are briefly analyzed.

Due to the vagueness of Chen's different structures (such as the reachable subsets from spans and upper bounds), we do not attempt to give further analysis of the complexity of Algorithm~3 beyond the work in Chen's paper~\cite{chen:2-maxsat-solved-in-p-time} for concern that we will not implement their algorithms as intended. Despite this, there still seem to be flaws and overgeneralizations in Chen's analysis.

When analyzing the runtime of the initial structure constructions and search algorithms, Chen separates the problem into three main subcategories, $\tau_1$, $\tau_2$ (which is further divided into $\tau_{21}$, $\tau_{22}$, and $\tau_{23}$), and $\tau_3$. While each subsection represents the runtime for a specific step or algorithm in Chen's solution, these runtimes are merely listed with minimal explanation~\cite{chen:2-maxsat-solved-in-p-time}. We believe further analysis should be used here to improve comprehension. For example, Chen states that the complexity for sorting the variable sequences in the $D_i$'s is clearly $O(nm\log m)$ without any further reasoning~\cite{chen:2-maxsat-solved-in-p-time}. Instead, it would be helpful to reiterate that this is because there are $n$ different $D_i$ sequences to sort that each contain $m$ variables. Similarly, for the complexity of constructing the $n$ $p$*-graphs, it is claimed that $\tau_{22}$ is $O(nm^2)$ since at most $O(m^2)$ time is needed to find the transitive closure over each graph's spans without explanation~\cite{chen:2-maxsat-solved-in-p-time}. This is because in the worst case each $p$-graph can have $m-2$ spans (besides the edges that Chen calls the ``main path'' of a $p$-graph), and therefore at most $(m-2) + (m-3) + \dots + 1 = \frac{(m-2)(m-1)}{2}$ edges will be created. 

Finally, Chen establishes a recursive relationship to analyze $\tau_3$, the average complexity of searching $G$ with the improved Algorithm~3. Since Chen claims that their analysis gives the average time complexity of $\tau_3$~\cite{chen:2-maxsat-solved-in-p-time}, they are in fact not providing the worst-case analysis of the algorithm and are incorrectly using Big-Oh notation. However, it is unclear if this mistake in notation was an error in wording (and Chen meant to say worst-case) or practice (and Chen is really giving an average time analysis).  

Regardless of Chen's incorrect usage of Big-Oh notation, their proof for the average complexity of $\tau_3$ still contains major concerns. For instance, they introduce constants that can take on undefined values (such as using $d$, which can take on the value of one and cause many logarithms in Chen's analysis to be undefined), and constants that lack definitions whatsoever (such as the undefined constant used to establish the base case of their recursive relation)~\cite{chen:2-maxsat-solved-in-p-time}. Furthermore, it is unclear how $d^{\lceil \log_d l \rceil} \leq l$ is derived in the computations to simplify Chen's Equation~6~\cite{chen:2-maxsat-solved-in-p-time}, and why the ceiling function is not applied to the exponent of $(log_d l )$. Despite their initial analysis, Chen notes that their recursive relation does not show that their proposed algorithm has a polynomially-bounded runtime, so the author attempts to give a lower bound.

In the final paragraph of Chen's Section~5~\cite{chen:2-maxsat-solved-in-p-time}, it is claimed that each branching node can be involved in at most $O(n)$ recursive calls. It is unclear how this claim is being proved, however it seems that Chen claims the number of recursive calls a node is involved in is at most $n-2$, but if there is an upper bound to that node it may be involved in an additional recursive call. Since the number of branching nodes is at most the number of nodes in a trie-like subgraph ($O(nm)$) and each recursive call is claimed to run in $O(nm^2)$, this would cause the worst case runtime of the algorithm to be $O(n \times nm \times nm^2) = O(n^3m^3)$. Furthermore, recall that since $n$ and $m$ were found to refer to the number of clauses and variables in the 2-DNF formula D, this paper (in Section~\ref{s:chens-algo-two}) defined $n_0$ and $m_0$ to be the respective number of clauses and variables in the corresponding 2-CNF formula. It is clear that in the worst-case $n = 2n_0$ as Chen converts every clause in the 2-CNF boolean formula into two new clauses in their DNF formula and similarly $m = n_0 + m_0$ as a new variable is created in the 2-DNF formula for every clause in the 2-CNF formula. Additionally, it is clear that in the worst-case there can be at most $2n_0$ unique variables in a 2-CNF formula if each clause contains two unique variables. Therefore, in the worst-case, it follows that
\begin{equation*}
n^3m^3 \leq 
(2n_0)^3(n_0 + m_0)^3 \leq 
(2n_0)^3(3n_0)^3 
= 216n_0^6.
\end{equation*}
Thus the worst-case runtime of Chen's solution in terms of the number of clauses in a 2-CNF formula is $O(n_0^6)$.

While analyzing Chen's work we could not find any specific algorithm that stood out as concretely running in exponential time. Nonetheless, we have provided reasons to doubt the validity of Chen's claims that their algorithm has a polynomial runtime. Furthermore, as we have shown through our counterexamples, regardless of whether Chen's Algorithm~3 does have a polynomial runtime, the algorithm does not always provide valid solutions for the 2-MAXSAT problem. 

\section{Chen's Use of Examples Over Proofs}
\label{s:examples-over-proofs}

Throughout the technical-report version~\cite{chen:2-maxsat-solved-in-p-time} and conference version~\cite{chen:2-maxsat-solved-in-p-time-conf} of Chen's papers, the 2-CNF boolean formula
\begin{equation*}
    C = (c_1 \vee c_2) \wedge (c_2 \vee \neg c_3) \wedge (c_3 \vee \neg c_1)
\end{equation*}
is used as an example to detail the steps of Chen's proposed solution to 2-MAXSAT and to show that their solution is correct, in place of formal definitions. In general, showing that an algorithm finds a solution to a single example is not a sufficient proof of correctness. Furthermore, some steps of Chen's solution such as the global ordering of the variable sequences and specific implementations of the $p$-graphs, $p$*-graphs, trie-like graph, and layered graph are only partially formalized. This is problematic as different design choices to implement these steps could lead to major variations in both the runtime and validity of Chen's proposed solution.

Chen's use of formula $C$ as a reoccurring example is also problematic as this formula is satisfiable~\cite{chen:2-maxsat-solved-in-p-time}. It is well known that the problem of deciding whether a 2-CNF boolean formula has at least one satisfying assignment (also called 2-SAT) is in $\p$. Therefore, there clearly exists a polynomial-time algorithm that can determine if $C$ has an assignment that satisfies at least $k$ clauses (with $1 \leq k \leq n$). Thus the 2-MAXSAT problem for $C$ (or any formula in 2-SAT) can be solved trivially, with the maximum number of satisfied clauses equal to the number of clauses in $C$. This leads us to further doubt Chen's algorithm, as the formula $C$ that they use to show the correctness and polynomial runtime of their algorithm can already be computed, in polynomial time, to have a maximum of three satisfied clauses. 

Finally, Chen fails to provide valid proofs that any of their proposed algorithms always return a correct result.
Indeed, for Algorithms~1 and 2, which we have shown to produce incorrect results on certain inputs, a proof of correctness cannot exist.
Chen's Proposition~2 claims to provide such a proof for Algorithms~1~and~2~\cite{chen:2-maxsat-solved-in-p-time}, but it assumes both that there are no issues in constructing the trie-like graph $G$, and that Algorithm~1 and Algorithm~2 will each work as informally described. A similarly-styled proof is also used for Chen's Proposition~3~\cite{chen:2-maxsat-solved-in-p-time}, which claims to show the correctness of Algorithm~3.

\section{Conclusion}

In this paper, we have provided examples where Chen's proposed algorithms fail to solve instances of the 2-MAXSAT problem. This disproves the central claims of both Chen's conference version~\cite{chen:2-maxsat-solved-in-p-time-conf} and technical-report version~\cite{chen:2-maxsat-solved-in-p-time}. Additionally, we have shown that due to poor definitions and vague analyses, Chen fails to establish whether the proposed algorithms run in polynomial time. We conclude that Chen has failed to prove the claim made in the title of their papers~\cite{chen:2-maxsat-solved-in-p-time-conf, chen:2-maxsat-solved-in-p-time}, and so has also failed to prove that $\p = \np$.

\paragraph{Acknowledgments}
We would like to thank
Michael~C.~Chavrimootoo,
Lane~A.~Hemaspaandra,
Harry~Liuson,
and
Zeyu~Nie
for their helpful comments on prior drafts. The authors are responsible for any remaining errors.

\bibliographystyle{alpha}
\bibliography{gry-reu,local_refs}

\end{document}